\documentclass[a4paper,11pt]{article}
\usepackage{pos}

\title{Exploring the Potential of the Pulsed Laser onboard the CALIPSO Satellite to Improve Calibration with VERITAS}
 \ShortTitle{Exploring the Potential of the CALIPSO Satellite to Improve Calibration with VERITAS}

\author*[a]{Gregory Foote}

\affiliation[a]{Department of Physics and Astronomy, University of Delaware, Newark, USA}

\onbehalf{for the VERITAS collaboration} 

\emailAdd{gregoryf@udel.edu}

\abstract{Imaging Atmospheric Cherenkov Telescopes (IACTs) are used to detect bright nanosecond-duration flashes of optical light originating from interactions of cosmic/gamma-rays in the atmosphere. A natural calibration source with similar characteristics does not exist; however, satellite-based laser systems provide a potential alternative. The CALIPSO satellite is one such facility which uses a suite of instruments to gather information about the atmosphere. Of particular interest is the CALIOP instrument, which emits 20-nanosecond laser pulses at 1064 nm and 532 nm at a rate of 20 Hz towards the Earth. The TAIGA-HiSCORE collaboration announced a detection of CALIOP laser pulses at the 37th ICRC in 2021, demonstrating that the laser footprint extends to at least tens of kilometers from the subsatellite point. We have used the VERITAS IACT to observe CALIPSO, and show here the results of using these observations to help to calibrate the array. We also discuss the potential of this technique for cross-calibration between different IACT facilities and for relative calibration between the telescopes of future large arrays.}

\ConferenceLogo{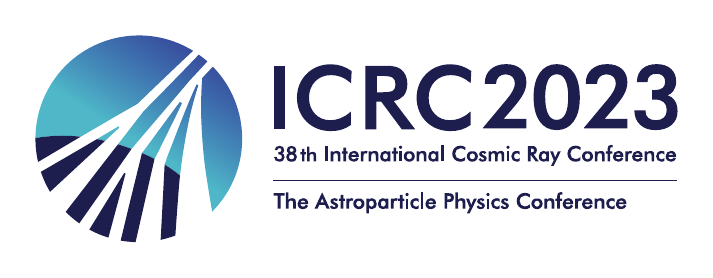}

\FullConference{%
38th International Cosmic Ray Conference (ICRC2023)\\
  26 July - 3 August, 2023\\
  Nagoya, Japan}

\begin{document}
\maketitle

\section{Introduction}
The Very Energetic Radiation Imaging Telescope Array System (VERITAS) is a set of four imaging atmospheric Cherenkov telescopes located at the Fred Lawrence Whipple Observatory (FLWO) in southern Arizona (31$\deg$ 40$'$N, 110$\deg$ 57$'$W,  1.3km a.s.l.), and is designed to observe the phenomenon of ``Cherenkov showers'' where large bursts of Cherenkov light are produced high in the atmosphere originating from the interactions between high energy gamma rays/cosmic rays and the Earth's atmosphere. Each telescope is a 12 m Davies-Cotton telescope, with all four telescopes roughly 100 m apart in a rectangular layout. Each 12 m reflector is made of 350 hexagonal mirrors which focus light onto a camera composed of an array of 499 photo-multiplier tubes (PMTs). Each PMT is fitted with a hexagonal light cone in order to remove the gaps between PMTs to tile the camera plane \citep{2011ICRC...12..137H}. In order to prevent being overwhelmed with data and to only collect images of Cherenkov showers, VERITAS employs a three-tiered trigger system. This trigger system begins with a pixel-level trigger associated with the signal measured by a single PMT exceeding a set threshold. The next level up is a telescope-level trigger associated with a group of three or more neighboring pixel triggers within 5 ns of each other in the same camera. Finally, an array-level trigger is formed if two or more telescope-level triggers enter a triggered state within 50 ns of each other after correcting for delays of the Cherenkov light between telescopes. Triggered events are read out using flash-ADCs which record 32-ns signal traces for each PMT with 2-ns samples \citep{2008ICRC....3.1539W}. 

The Cloud-Aerosol LIDAR and Infrared Pathfinder Satellite Observations (CALIPSO) satellite \citep{2009JAtOT..26.2310W} is a joint atmospheric science mission built, maintained, and launched by NASA and CNES, which operates at an altitude of 700 km. Its primary instrument, the Cloud-Aerosol LIdar with Orthogonal Polarization (CALIOP) instrument, is of particular interest to this study because it hosts two lasers: a 532 nm laser and 1064 nm laser, both of which produce 20-ns pulses at a rate of 20.16 Hz directed towards the Earth . While 532 nm is within the detectable wavelength range of Cherenkov telescopes, the footprint on the ground was expected to be too small (on the order of 100 m) to be observable from anywhere but the direct satellite path. However, the TAIGA-HiScore telescope announced in 2021 that they were able to detect CALIPSO \citep{2022icrc.confE.876P}, which refuted the assumption about the footprint and spurred VERITAS to begin observations. Using VERITAS, CALIPSO has been shown to be consistently detectable. We have conducted an observation campaign of CALIPSO passes, as well as developed analysis methods to exploit its potential as a calibration source for Cherenkov telescopes. Potentially visible CALIPSO passes occur approximately every three days and, as a stable, nanosecond-pulsed light source located at large distance, its potential as a calibration source is unique.

\section{VERITAS Observations of CALIPSO}
The observation campaign of CALIPSO with VERITAS covers a wide range of azimuth angles, and elevation angles between zenith and $55\deg$ elevation. As the dataset of CALIPSO observations grows, we are developing an understanding of the correlation between the elevation and azimuth of the satellite and VERITAS's response. This allows us to predict the response at a given sky location. This also provides an opportunity to study the laser footprint of CALIPSO which could potentially help other observatories utilize CALIPSO for internal or cross-calibrations, including with VERITAS.

\begin{figure}
    \centering
    \includegraphics[width=1\linewidth]{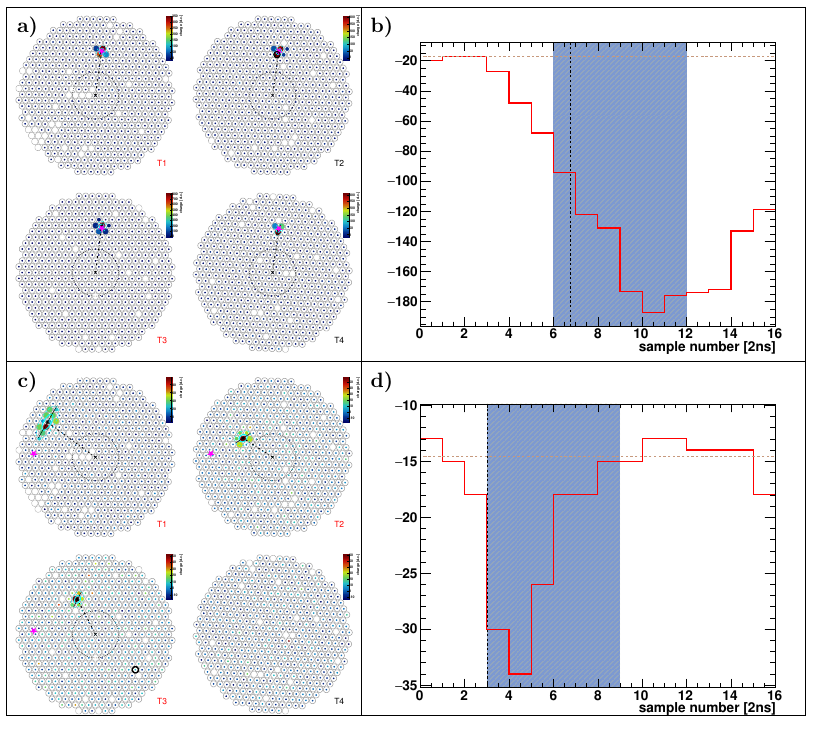}
    \caption{Two events as seen by VERITAS are showcased, the top row produced by CALIPSO, and the bottom row produced by a regular cosmic-ray, \textbf{a)} shows a CALIPSO pulse as seen by each telescope, \textbf{b)} shows digital counts versus time from the bright pixel in telescope 4 for this pulse, \textbf{c)} shows a high-altitude cosmic-ray as seen by each telescope, and \textbf{d)} shows digital counts versus time from the bright pixel in telescope 3 for this cosmic-ray. The typical bounds of the pulse sum are shown with the blue boxes in \textbf{c)} and \textbf{d)}.}
    \label{fig:calipso_vs_cosmic_ray}
\end{figure}

We identify CALIPSO pulses in our data by exploiting two characteristics: the lack of parallax and the pulse arrival time. First, we require the centers of the images from each telescope (within a single event) to be within two pixel widths ($0.3\deg$) of each other; as seen in figure \ref{fig:calipso_vs_cosmic_ray}. We also calculate the phase of the pulse arrival time with respect to the 20.16 Hz laser pulse frequency. If a large collection of events with phases within 0.007 of each other appears in the observation run, they are classed as CALIPSO pulses. Once the pulses are identified in our data, we perform analysis on them to get calibration products.

\section{Camera Rotations}
During the construction of VERITAS, the cameras were placed on their support structures with a slight rotation accidentally. It is therefore necessary to measure and correct for this effect in our analysis. Initially this rotation was measured using a plumb bob, but the measurement became simpler once CCD cameras were installed which record images of both the PMT camera of each telescope and the sky in the direction to which they are pointed \citep{2008AIPC.1085..657H}. The current method positions the telescopes at $180\deg$ azimuth and angled towards the Celestial Equator, such that a bright star travels through the middle of all the cameras. When the star travels through the field-of-view, the track across the middle of the camera is compared between the CCD camera's view and each individual telescope to determine the amount of rotation which is needed to correct for the camera rotations. CALIPSO pulses can be used to perform a similar measurement, because we know that the position and track of CALIPSO should be the same in every camera. We can also potentially measure changes in camera rotations at different elevations and across time, if such changes exist. 

\begin{figure}
    \centering
    \includegraphics[width=1\linewidth]{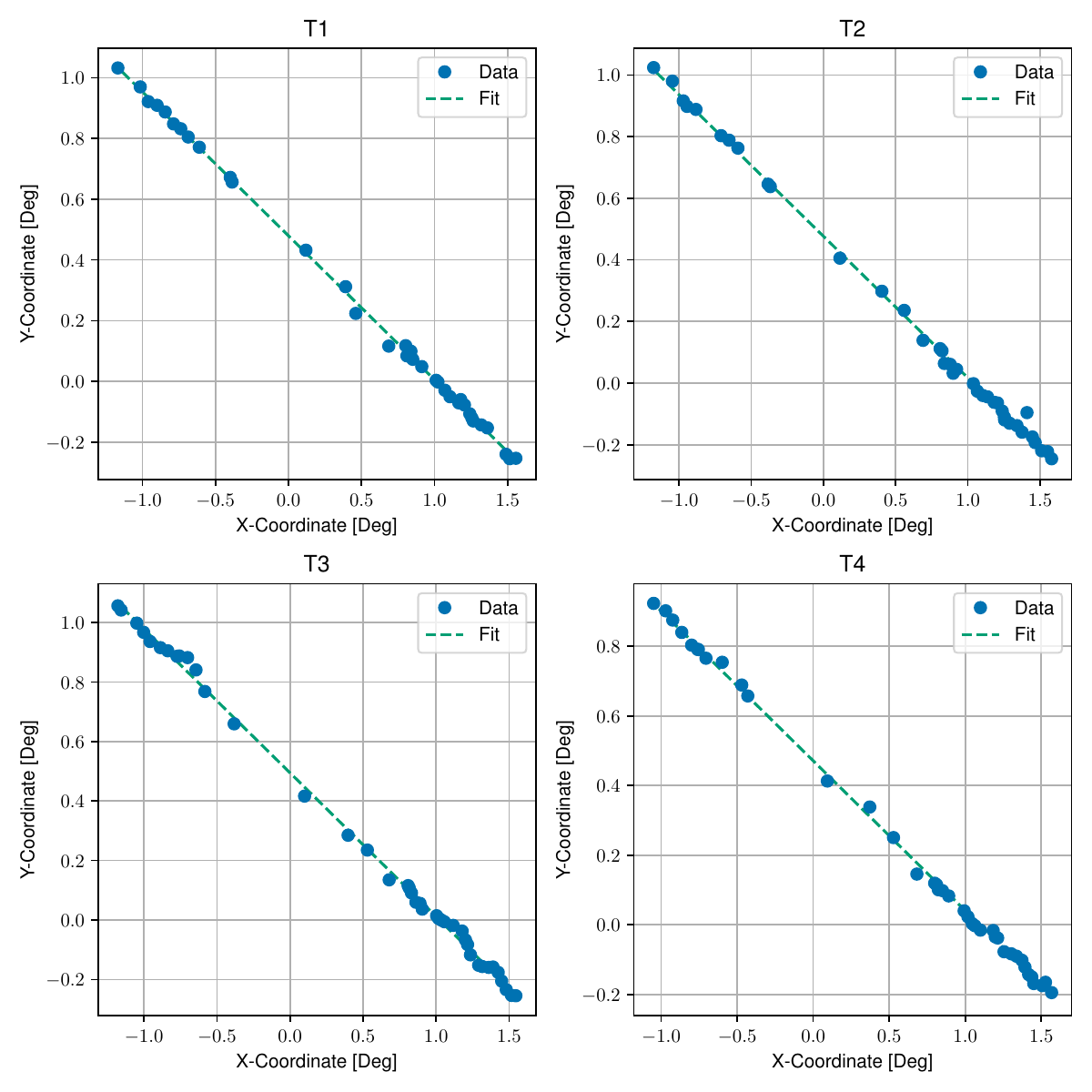}
    \caption{The fitted path through the VERITAS camera (dotted line) along with each image center for every captured event (blue dots).}
    \label{fig:telescope_v_path}
\end{figure}

Using data from an observation of CALIPSO taken on January 29$^{\text{th}}$, 2023 at $58\deg$ elevation, we calculated the camera rotations by first solving for the angle between the positive X-axis and CALIPSO's path in each camera. This angle is found by first performing a least-squares fit of a line on the path CALIPSO makes in the camera, as seen in figure \ref{fig:telescope_v_path}. We then use equation \ref{eq:arctan_slope} to find the angle and equation \ref{eq:arctan_slope_error} to find its error.
\begin{equation} \label{eq:arctan_slope}
\theta_i = \arctan\left(m_{i\text{,fit}}\right)
\end{equation}

\begin{equation} \label{eq:arctan_slope_error}
\delta\theta_i = \left|\arctan\left(m_{i\text{,fit}}\right)\right|\cdot\left|\frac{\delta m_{i\text{,fit}}}{m^3_{i\text{,fit}} + m_{i\text{,fit}}}\right|
\end{equation}

The relative rotation of each telescope is solved for by setting telescope 1 to zero. Doing this provides a comparison to the currently used values within VERITAS, which used the traditional method. The comparison of this analysis and the currently used values is found in Table \ref{table:rotation_results}.

\begin{table}
\centering
\begin{tabular}{|c|c|c|c|c|c|} 
\hline Telescope ID & Current Rotations & This Analysis \\ 
\hline 1 & $0.$ & $0. \pm 0.077$ \\ 
\hline 2 & $0.612$ & $0.692 \pm 0.092$ \\ 
\hline 3 & $-0.806$ & $-0.426 \pm 0.088$ \\ 
\hline 4 & $2.027$ & $2.086 \pm 0.079$ \\ 
\hline 
\end{tabular}
\caption{Table of the measured rotations from this analysis and the rotation measurements currently used in VERITAS. The rotations have been modified from their original values to be relative to Telescope 1 for ease of comparison; all measurements are in degrees.}
\label{table:rotation_results}
\end{table}

\section{Relative Throughputs}
The optical systems of VERITAS experience significant change over time. The telescope mirrors suffer from degradation due to exposure to the southern Arizona desert and climate throughout the year. These changes are corrected for with maintenance tasks including mirror cleanings, swaps, replacements and re-coatings. Alongside this, the PMTs also degrade over time from two effects: operating at high voltages over long periods of time and charged positive ions accumulating within them. Measuring these effects is crucial in maintaining an accurate model of VERITAS within our simulations. A full discussion of current VERITAS calibration techniques to measure the impacts on the throughputs of a given telescope (defined as the product of the reflectivity, accounting for the mirrors, and the gain, accounting for the PMTs) and the results thereof can be found in \cite{2022A&A...658A..83A}.

We are exploring the possibility of utilizing CALIPSO for performing some aspects of this calibration. Performing this calibration with CALIPSO is possible because the spacing of the VERITAS telescopes relative to the altitude of CALIPSO, 100 meters versus 700 kilometers, ensures that the amount of energy received by each telescope should be approximately the same.

The total amount of light captured by the VERITAS camera is estimated to be proportional to the sum of all traces from PMTs which pass standard image cleaning procedures. The portion of the trace which is included in this sum normally begins roughly halfway through the rising edge of the captured pulse and covers 12 ns \citep{2008ICRC....3.1325D}. This method, called the double-pass method, works well for the pulses produced from Cherenkov showers, which are usually entirely captured within the 32-ns FADC window stored. However, CALIPSO pulses are relatively slow, with a full-width at half-maximum of 20 ns, making this method unreliable. The differences in the pulse characteristics are shown in figure \ref{fig:calipso_vs_cosmic_ray}. For this analysis, the analysis method was adjusted to begin the pulse sum at the start of the 32 ns captured window and end at the maximum of the pulse. While normally this would significantly increase the amount of noise, the uniformity and intensity of the captured CALIPSO pulses mitigates this.

\begin{figure}
    \centering
    \includegraphics[width=1\linewidth]{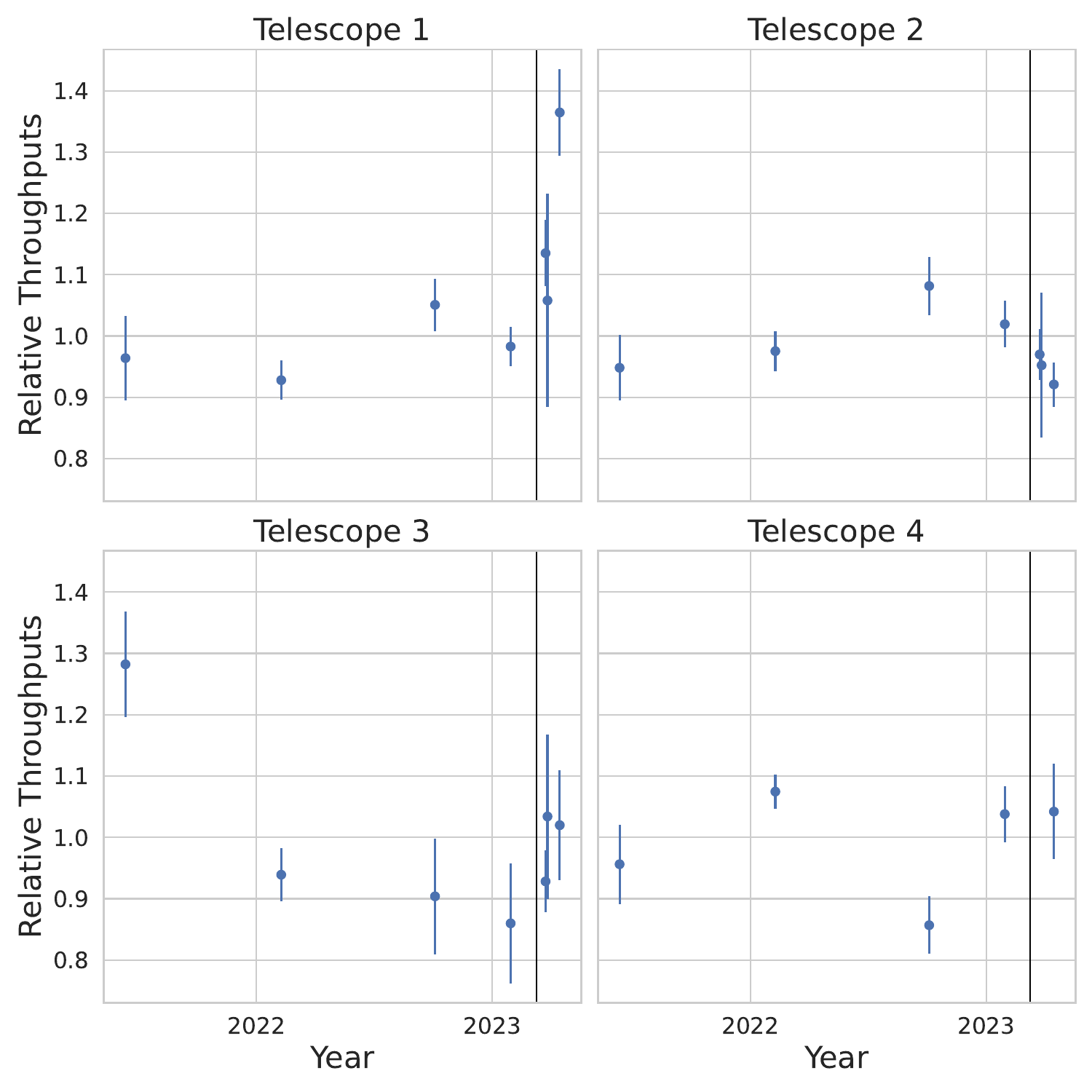}
    \caption{The relative throughputs for each telescope across time, calculated using CALIPSO laser pulses. The completion date of mirror facet replacements and alignment on Telescope 1, March 10$^{\text{th}}$, 2023, is indicated by the solid black line.}
    \label{fig:rel_mean_w_err}
\end{figure}

The above method yields the amount of light captured per pulse per telescope; to provide a more useful comparison we find the average of the light intensity across all pulses for a given telescope in a single transit; along with the error of that average. We move to a per-transit basis as opposed to working with individual pulses because across the entire transit the laser does not significantly change. To get an individual telescope's relative throughput for a transit, we divide the response from each telescope by the error-weighted average of all telescopes for that transit. These relative throughputs can then be compared across different transits at different elevations. The results from 7 CALIPSO transits are shown in figure \ref{fig:rel_mean_w_err}. An early verification of our method is to use it to show an increase of the relative throughput of Telescope 1 after March 10$^{\text{th}}$, 2023. Between March 1$^{\text{st}}$ and 10$^{\text{th}}$, 91 mirror facets were replaced and aligned which increased the reflectivity of Telescope 1. While more data is required to understand how well this method performs relative to traditional methods, the method shows initial promise.

\section{Conclusion}
CALIPSO represents an unique calibration tool as it guarantees a point-like, bright, pulsed, distant optical source whose location and energy are known. This enables several calibrations of VERITAS, two of which are explored here, with data taken under normal operating conditions. Additional calibration uses of CALIPSO which we have not yet explored include measurements of the telescope optical point spread functions and atmospheric model validation, as well as studies of the stability of the telescope properties over time. As the dataset of CALIPSO observations with VERITAS grows, our understanding of the VERITAS response to the laser will improve, which will help with current and future calibration studies. 

\section{Acknowledgements}

This research is supported by grants from the U.S. Department of Energy Office of Science, the U.S. National Science Foundation and the Smithsonian Institution, by NSERC in Canada, and by the Helmholtz Association in Germany. This research used resources provided by the Open Science Grid, which is supported by the National Science Foundation and the U.S. Department of Energy's Office of Science, and resources of the National Energy Research Scientific Computing Center (NERSC), a U.S. Department of Energy Office of Science User Facility operated under Contract No. DE-AC02-05CH11231. We acknowledge the excellent work of the technical support staff at the Fred Lawrence Whipple Observatory and at the collaborating institutions in the construction and operation of the instrument.

\bibliographystyle{JHEP}
\bibliography{bibl}
\clearpage

\section*{Full Author List: VERITAS Collaboration}

\scriptsize
\noindent
A.~Acharyya$^{1}$,
C.~B.~Adams$^{2}$,
A.~Archer$^{3}$,
P.~Bangale$^{4}$,
J.~T.~Bartkoske$^{5}$,
P.~Batista$^{6}$,
W.~Benbow$^{7}$,
J.~L.~Christiansen$^{8}$,
A.~J.~Chromey$^{7}$,
A.~Duerr$^{5}$,
M.~Errando$^{9}$,
Q.~Feng$^{7}$,
G.~M.~Foote$^{4}$,
L.~Fortson$^{10}$,
A.~Furniss$^{11, 12}$,
W.~Hanlon$^{7}$,
O.~Hervet$^{12}$,
C.~E.~Hinrichs$^{7,13}$,
J.~Hoang$^{12}$,
J.~Holder$^{4}$,
Z.~Hughes$^{9}$,
T.~B.~Humensky$^{14,15}$,
W.~Jin$^{1}$,
M.~N.~Johnson$^{12}$,
M.~Kertzman$^{3}$,
M.~Kherlakian$^{6}$,
D.~Kieda$^{5}$,
T.~K.~Kleiner$^{6}$,
N.~Korzoun$^{4}$,
S.~Kumar$^{14}$,
M.~J.~Lang$^{16}$,
M.~Lundy$^{17}$,
G.~Maier$^{6}$,
C.~E~McGrath$^{18}$,
M.~J.~Millard$^{19}$,
C.~L.~Mooney$^{4}$,
P.~Moriarty$^{16}$,
R.~Mukherjee$^{20}$,
S.~O'Brien$^{17,21}$,
R.~A.~Ong$^{22}$,
N.~Park$^{23}$,
C.~Poggemann$^{8}$,
M.~Pohl$^{24,6}$,
E.~Pueschel$^{6}$,
J.~Quinn$^{18}$,
P.~L.~Rabinowitz$^{9}$,
K.~Ragan$^{17}$,
P.~T.~Reynolds$^{25}$,
D.~Ribeiro$^{10}$,
E.~Roache$^{7}$,
J.~L.~Ryan$^{22}$,
I.~Sadeh$^{6}$,
L.~Saha$^{7}$,
M.~Santander$^{1}$,
G.~H.~Sembroski$^{26}$,
R.~Shang$^{20}$,
M.~Splettstoesser$^{12}$,
A.~K.~Talluri$^{10}$,
J.~V.~Tucci$^{27}$,
V.~V.~Vassiliev$^{22}$,
A.~Weinstein$^{28}$,
D.~A.~Williams$^{12}$,
S.~L.~Wong$^{17}$,
and
J.~Woo$^{29}$\\
\\
\noindent
$^{1}${Department of Physics and Astronomy, University of Alabama, Tuscaloosa, AL 35487, USA}

\noindent
$^{2}${Physics Department, Columbia University, New York, NY 10027, USA}

\noindent
$^{3}${Department of Physics and Astronomy, DePauw University, Greencastle, IN 46135-0037, USA}

\noindent
$^{4}${Department of Physics and Astronomy and the Bartol Research Institute, University of Delaware, Newark, DE 19716, USA}

\noindent
$^{5}${Department of Physics and Astronomy, University of Utah, Salt Lake City, UT 84112, USA}

\noindent
$^{6}${DESY, Platanenallee 6, 15738 Zeuthen, Germany}

\noindent
$^{7}${Center for Astrophysics $|$ Harvard \& Smithsonian, Cambridge, MA 02138, USA}

\noindent
$^{8}${Physics Department, California Polytechnic State University, San Luis Obispo, CA 94307, USA}

\noindent
$^{9}${Department of Physics, Washington University, St. Louis, MO 63130, USA}

\noindent
$^{10}${School of Physics and Astronomy, University of Minnesota, Minneapolis, MN 55455, USA}

\noindent
$^{11}${Department of Physics, California State University - East Bay, Hayward, CA 94542, USA}

\noindent
$^{12}${Santa Cruz Institute for Particle Physics and Department of Physics, University of California, Santa Cruz, CA 95064, USA}

\noindent
$^{13}${Department of Physics and Astronomy, Dartmouth College, 6127 Wilder Laboratory, Hanover, NH 03755 USA}

\noindent
$^{14}${Department of Physics, University of Maryland, College Park, MD, USA }

\noindent
$^{15}${NASA GSFC, Greenbelt, MD 20771, USA}

\noindent
$^{16}${School of Natural Sciences, University of Galway, University Road, Galway, H91 TK33, Ireland}

\noindent
$^{17}${Physics Department, McGill University, Montreal, QC H3A 2T8, Canada}

\noindent
$^{18}${School of Physics, University College Dublin, Belfield, Dublin 4, Ireland}

\noindent
$^{19}${Department of Physics and Astronomy, University of Iowa, Van Allen Hall, Iowa City, IA 52242, USA}

\noindent
$^{20}${Department of Physics and Astronomy, Barnard College, Columbia University, NY 10027, USA}

\noindent
$^{21}${ Arthur B. McDonald Canadian Astroparticle Physics Research Institute, 64 Bader Lane, Queen's University, Kingston, ON Canada, K7L 3N6}

\noindent
$^{22}${Department of Physics and Astronomy, University of California, Los Angeles, CA 90095, USA}

\noindent
$^{23}${Department of Physics, Engineering Physics and Astronomy, Queen's University, Kingston, ON K7L 3N6, Canada}

\noindent
$^{24}${Institute of Physics and Astronomy, University of Potsdam, 14476 Potsdam-Golm, Germany}

\noindent
$^{25}${Department of Physical Sciences, Munster Technological University, Bishopstown, Cork, T12 P928, Ireland}

\noindent
$^{26}${Department of Physics and Astronomy, Purdue University, West Lafayette, IN 47907, USA}

\noindent
$^{27}${Department of Physics, Indiana University-Purdue University Indianapolis, Indianapolis, IN 46202, USA}

\noindent
$^{28}${Department of Physics and Astronomy, Iowa State University, Ames, IA 50011, USA}

\noindent
$^{29}${Columbia Astrophysics Laboratory, Columbia University, New York, NY 10027, USA}

\end{document}